\documentclass[12pt]{article}
\usepackage{axodraw,bbold}

\catcode`@=12
\begin{document}
\thispagestyle{empty}
\begin{flushright} 
UCRHEP-T424\\ 
November 2006\
\end{flushright}
\vspace{0.5in}
\begin{center}
{\LARGE	\bf Neutrino Mass, Dark Matter,\\ and Leptogenesis\\}
\vspace{1.5in}
{\bf Ernest Ma\\}
\vspace{0.2in}
{\sl Physics and Astronomy Department, University of California, Riverside, 
California 92521 \\}
\vspace{1.5in}
\end{center}

\begin{abstract}\
It is proposed that dark matter is the origin of neutrino mass, thereby 
linking inexorably two undisputed (and seemingly unrelated) pieces of 
evidence for physics beyond the Standard Model.  Leptogenesis at the 
TeV scale may also be possible, as well as a measurable contribution 
to the muon anomalous magnetic moment.
\end{abstract}

\vskip 2cm

\noindent $^*$Talk at NOW2006, Conca Specchiulla, Otranto, Italy.

\newpage
\baselineskip 24pt

\section{INTRODUCTION}

Any theory beyond the Standard Model (SM) should incorporate neutrino mass and 
dark matter.  Are they related?  In this talk, I propose that neutrino mass 
is due to the existence of dark matter.  I will discuss some recent models 
and their phenomenological consequences.

A candidate for dark matter should be neutral and stable, the latter implying 
at least an exactly conserved odd-even ($Z_2$) symmetry.  In the Minimal 
Supersymmetric Standard Model (MSSM), the lightest neutral particle having 
odd $R$ parity is a candidate.  It is usually assumed to be a fermion, i.e. 
the lightest neutralino. [The lightest neutral boson, presumably a scalar 
neutrino, is ruled out phenomenologically.]

If all one wants is dark matter, the simplest way is to add a second Higgs 
double $(\eta^+,\eta^0)$ \cite{bhr06} which is odd under $Z_2$ with all SM 
particles even.  This differs from the scalar MSSM $(\tilde \nu, \tilde l)$ 
doublet, because $\eta^0_R$ and $\eta^0_I$ are split in mass by the $Z_2$ 
conserving term $(\lambda_5/2)(\Phi^\dagger \eta)^2 + H.c.$ which is absent 
in the MSSM.

\section{NEUTRINO MASS AND DARK MATTER}

To obtain a small Majorana neutrino mass in the SM, consider the unique 
dimension-five operator \cite{w79}
\begin{equation}
{f_{\alpha \beta} \over 2 \Lambda} (\nu_\alpha \phi^0 - l_\alpha \phi^+)
(\nu_\beta \phi^0 - l_\beta \phi^+).
\end{equation}
It has exactly 3 tree-level realizations \cite{m98} by inserting the 
appropriate intermediate state: (I) $N$, (II) 
$(\xi^{++},\xi^+,\xi^0)$, (III) $(\Sigma^+,\Sigma^0,\Sigma^-)$; and 3 
generic one-loop realizations \cite{m98}: (IV) one external scalar line 
coupled to the internal fermion line and the other to the internal scalar 
line, (V) both coupled to the scalar line, (VI) both coupled to the 
fermion line.

Radiative mechanism (IV) dominates the literature, the prime example being 
the Zee model \cite{z80}.  I propose instead \cite{m06} mechanism (V) with 
the addition of $N_i,~i=1,2,3$ and $(\eta^+,\eta^0)$, all being odd under 
$Z_2$.  As a result, either $\eta^0_R$ or $\eta^0_I$ is dark matter with 
mass 60 to 80 GeV \cite{bhr06}, or the lightest $N_i$ is dark matter, 
with all masses of order 350 GeV or less \cite{kms06} for the new particles.

\section{NEUTRINO MASS AND LEPTOGENESIS}

Because of the assumed $Z_2$ symmetry, $\nu_\alpha$ does not couple to $N_i$ 
through $\phi^0$.  Hence $N_i$ is not the Dirac mass partner of $\nu_\alpha$ 
as in the canonical seesaw model.  Instead $\nu_\alpha$ couples to $N_i$ 
through $\eta^0$ (which has no vacuum expectation value) and obtains a 
radiative Majorana mass in one loop, i.e. 
\begin{equation}
({\cal M}_\nu)_{\alpha \beta} = \sum_i {h_{\alpha i} h_{\beta i} M_i \over 
16 \pi^2} \left[ f \left({M_i^2 \over m_R^2} \right) - f \left({M_i^2 \over 
m_I^2} \right) \right],
\end{equation}
where $f(x)=-\ln x/(1-x)$. Let $m_R^2-m_I^2 = 2 \lambda_5^2 << m_0^2 = (m_R^2+ 
m_I^2)/2$, then
\begin{equation}
({\cal M}_\nu)_{\alpha \beta} = \sum_i {h_{\alpha i} h_{\beta i} \over M_i} 
I \left( {M_i^2 \over m_0^2} \right),
\end{equation}
where
\begin{equation}
I(x) = {\lambda_5 v^2 \over 8 \pi^2} \left( {x \over 1-x} \right) \left[ 
1 + {x \ln x \over 1-x} \right].
\end{equation}
For $x_i >> 1$, i.e. $N_i$ very heavy,
\begin{equation}
({\cal M}_\nu)_{\alpha \beta} = {\lambda_5 v^2 \over 8 \pi^2} \sum_i 
{h_{\alpha i} h_{\beta i} \over M_i} [\ln x_i - 1]
\end{equation}
instead of the canonical seesaw expression of $v^2 \sum_i h_{\alpha i} h_{\beta i} 
/M_i$.  In leptogenesis, the lightest $M_i$ may then be much below the 
Davidson-Ibarra bound \cite{di02} of about $10^9$ GeV, thus avoiding a 
potential conflict of gravitino overproduction and thermal leptogenesis. 
In this scenario, $\eta^0$ is dark matter.

\section{MUON $g-2$ AND NEUTRINO MASS}

Another model of this class has recently been proposed \cite{hkmr06} again 
with an exactly conserved $Z_2$ where $N_i, N^c_i, (\eta^+,\eta^0), \chi^-$ 
are odd.  Lepton number, i.e. $U(1)_L$, is also assigned to these fields: 
$1$, $-1$, $0$, $0$ respectively.

\begin{figure}[htb]
\begin{center}\begin{picture}(250,80)(10,45)
\ArrowLine(40,50)(80,50)
\ArrowLine(120,50)(160,50)
\ArrowLine(120,50)(80,50)
\ArrowLine(200,50)(160,50)
\Text(60,35)[b]{$\mu$}
\Text(180,34)[b]{$\mu^c$}
\Text(100,32)[b]{$N_i^c$}
\Text(140,33)[b]{$N_i$}
%\Text(125,97)[b]{$\chi^-$}
\Text(75,70)[b]{$\eta^+$}
\Text(168,72)[b]{$\chi^-$}
%\Text(91,112)[b]{$\phi^0$}
%\Text(173,116)[b]{$\gamma$}
\Text(120,125)[b]{$\gamma$}
%\DashArrowLine(100,85)(95,111){3}
%\Photon(150,85)(165,111){2}{3}
\Photon(120,90)(120,120){2}{3}
\DashArrowArc(120,50)(40,90,180){3}
\DashArrowArcn(120,50)(40,90,0){3}
%\DashArrowArcn(120,50)(40,120,60){3}
\end{picture}
\end{center}
\caption[]{Dominant contributions to muon anomalous magnetic moment.}
\end{figure}
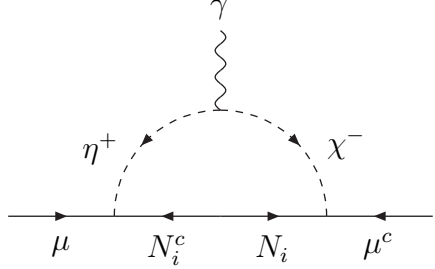

The muon gets an anomalous magnetic moment from the Dirac mass terms linking 
$N_i$ with $N^c_i$ and the mixing of $\eta^\pm$ with $\chi^\pm$ (forming the 
eigenstates $X$ and $Y$ with mixing angle $\theta$), resulting in
\begin{equation}
\Delta a_\mu = {-\sin \theta \cos \theta \over 16 \pi^2} \sum_i h_{\mu i} h'_{\mu i} 
{m_\mu \over M_i} [F(x_i)-F(y_i)],
\end{equation}
where $x_i = m_X^2/M_i^2$, $y_i = m_Y^2/M_i^2$, and $F(x)=[1-x^2+2x\ln x]
/(1-x)^3$.  Let $y_i << x_i \simeq 1$, $M_i \sim 1$ TeV, $(-h_{\mu i} 
h'_{\mu i} \sin \theta \cos \theta/24 \pi^2) \sim 10^{-5}$, then 
$\Delta a_\mu \sim 10^{-9}$, whereas $(\Delta a_\mu)_{\rm exp't} = 
(22.4 \pm 10)~{\rm to}~(26.1 \pm 9.4) \times 10^{-10}$.

\begin{figure}[htb]
\begin{center}\begin{picture}(250,80)(10,45)
\ArrowLine(40,50)(80,50)
\ArrowLine(120,50)(160,50)
\ArrowLine(120,50)(80,50)
\ArrowLine(200,50)(160,50)
\Text(60,35)[b]{$\nu_\alpha$}
\Text(180,34)[b]{$\nu_\beta$}
\Text(100,32)[b]{$N_i^c$}
\Text(140,31)[b]{$N_j^c$}
\Text(75,70)[b]{$\eta^0$}
\Text(168,72)[b]{$\eta^0$}
\Text(95,115)[b]{$\phi^0$}
\Text(150,115)[b]{$\phi^0$}
\DashArrowLine(100,110)(120,90){3}
\DashArrowLine(140,110)(120,90){3}
\DashArrowArc(120,50)(40,90,180){3}
\DashArrowArcn(120,50)(40,90,0){3}
\end{picture}
\end{center}
\caption[]{Radiative Majorana neutrino mass.}
\end{figure}
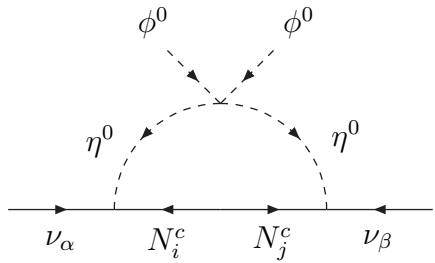

To obtain a nonzero neutrino mass, $U(1)_L$ is broken softly by the terms
\begin{equation}
{1 \over 2} m_{ij} N^c_i N^c_j + {1 \over 2} m'_{ij} N_i N_j + H.c.,
\end{equation}
resulting in the residual symmetry $(-1)^L$.  The one-loop radiative 
Majorana neutrino mass matrix is then given by
\begin{eqnarray}
({\cal M}_\nu)_{\alpha \beta} = \sum_{i,j} {h_{\alpha i} h_{\beta j} 
\lambda_5 v^2 m_{ij} \over 8 \pi^2 (M^2_i-M^2_j)} \left[ {M^2_i \over 
m_0^2-M_i^2} + {M_i^4 \ln (M_i^2/m_0^2) \over (m_0^2-M_i^2)^2} - (i 
\leftrightarrow j) \right].
\end{eqnarray}
Let $M_{i,j} \sim 1$ TeV, $m_{ij} \sim 0.1$ GeV, $h_{\alpha i} \sim 10^{-2}$, 
$\lambda_5 \sim 0.1$, $m_0 \sim v \sim 10^2$ GeV, then the entries of \
${\cal M}_\nu$ are of order 0.1 eV.

Suppose the $h_{\alpha 1}$ couplings are very small, i.e. $N^c_1$ decouples 
from ${\cal M}_\nu$, it is still possible to obtain a realistic neutrino 
mass matrix.  For example, let
\begin{equation}
h_{\alpha i} \simeq h \pmatrix{0 & 1 & 0 \cr 0 & 0 & 1/\sqrt 2 \cr 0 & 0 & 
1/\sqrt 2},
\end{equation}
then using the $[\nu_e, (\nu_\mu+\nu_\tau)/\sqrt 2, (-\nu_\mu+\nu_\tau)
/\sqrt 2]$ basis,
\begin{equation}
{\cal M}_\nu \simeq h^2 \pmatrix{\tilde m_{22} & \tilde m_{23} & 0 \cr 
\tilde m_{23} & \tilde m_{33} & 0 \cr 0 & 0 & 0},
\end{equation}
i.e. $\theta_{23} = \pi/4$, $\theta_{13} = 0$, $m_3=0$ (inverted ordering).

\section{NOVEL TEV LEPTOGENESIS}

Let $(N_1,N^c_1)$ be the lightest pair with $h_{\alpha 1},h'_{\alpha 1} 
\sim 10^{-7}$ to satisfy the out-of-equilibrium condition for leptogenesis 
at the TeV scale.  Rotate
\begin{equation}
\pmatrix{m'_{11} & M_1 \cr M_1 & m_{11}} \to \pmatrix{-M_1+A & B \cr B 
& M_1+A},
\end{equation}
where $A = (m'_{11}+m_{11})/2$, $B=(m'_{11}-m_{11})/2$.  Choose phases so that 
$M_1 > 0$, $A > 0$ are real and $B = |B|\exp(i \alpha)$.  Diagonalize 
above matrix with
\begin{equation}
\pmatrix{ \exp(i \beta) \cos \theta & -\sin \theta \cr \sin \theta & 
\exp(-i \beta) \cos \theta},
\end{equation}
then $\sin \beta = -M_1 \tan \alpha/C$, $\cos \beta = A/C$, $\tan 2 \theta = 
\cos \alpha |B| C/A M_1$, where $C = (A^2 + M_1^2 \tan^2 \alpha)^{1/2}$. 
Let $A^2 << M_1^2 \tan^2 \alpha$, then $\exp(i \beta) = (A/M_1 \tan \alpha) 
- i$, and the lepton asymmetry is given by
\begin{eqnarray}
\left( {-1 \over 64 \pi} \right) {|B|^2 \sin \alpha \cos \alpha \over 
A^2 + |B|^2 \sin^2 \alpha} \left[ {4(\sum_\alpha |h_{\alpha 1}|^2)^2 - 
(\sum_\alpha |h'_{\alpha 1}|^2 )^2 \over 2\sum_\alpha |h_{\alpha 1}|^2 
+ \sum_\alpha |h'_{\alpha 1}|^2} \right].
\end{eqnarray}
The novel feature of this mechanism is that $CP$ violation originates in the 
mass matrix, not the Yukawa couplings.  However, because $h_{\alpha 1}, 
h'_{\alpha 1} \sim 10^{-7}$, this effect is too small.  This means that 
$(N_2,N^c_2)$ must also be considered, and the complete expression of the 
asymmetry becomes rather complicated.  Now because $h'_{\alpha 2}$ are 
mostly unconstrained, this asymmetry may well be of order $10^{-6}$ for 
$M_{1,2}$ of order 1 TeV for a realistic scenario of leptogenesis, which is 
also verifiable at the forthcoming Large Hadron Collider (LHC).

\section{CONCLUSION}

The evidence of dark matter signals a new class of particles at the TeV scale, 
which may manifest themselves indirectly through loop effects.  They may be 
responsible for neutrino mass, muon anomalous magnetic moment, as well as 
leptogenesis.  Two simple examples predict observable bosonic dark matter 
at the electroweak scale, and perhaps also neutral singlet fermions at the 
TeV scale.

\section*{Afterword}

This talk was given on September 11, 2006.  Exactly 5 years ago, I was 
also in Italy giving a talk during that most fateful event of recent 
times.

\section*{Acknowledgement}

I thank Gianluigi Fogli, Carlo Giunti, and the other organizers of NOW2006 
for their great hospitality at Conca Specchiulla, Otranto. 
This work was supported in part by the U.~S.~Department of Energy under 
Grant No.~DE-FG03-94ER40837.

%\newpage

\end{document}